# Optical detection of long electron spin transport lengths in a monolayer semiconductor

L. Ren[1*], L. Lombez[1*], C. Robert[1*†], D. Beret[1], D. Lagarde[1], B. Urbaszek[1], P. Renucci[1], T. Taniguchi[2], K. Watanabe[3], S.A. Crooker[4], X. Marie[1†]

[1]Université de Toulouse, INSA-CNRS-UPS, LPCNO, 135 Av. Rangueil, 31077 Toulouse, France
[2]International Center for Materials Nanoarchitectonics, National Institute for Materials Science, 1-1 Namiki, Tsukuba 305-00044, Japan
[3]Research Center for Functional Materials, National Institute for Materials Science, 1-1 Namiki, Tsukuba 305-00044, Japan
[4]National High Magnetic Field Laboratory, Los Alamos National Laboratory, Los Alamos, New Mexico 87545, USA

*Using a spatially-resolved optical pump-probe experiment, we measure the lateral transport of spin/valley polarized electrons over very long distances (tens of micrometers) in a single WSe$_2$ monolayer. By locally pumping the Fermi sea of 2D electrons to a high degree of spin/valley polarization (up to 75%) using circularly-polarized light, the lateral diffusion of the electron polarization can be mapped out via the photoluminescence induced by a spatially-separated and linearly-polarized probe laser. Up to 25% spin-valley polarization is observed at pump-probe separations up to 20 microns. Characteristic spin/valley diffusion lengths of 18 ± 3 µm are revealed at low temperatures. The dependence on temperature, pump helicity, pump intensity, and electron density highlight the key roles played by spin relaxation time and pumping efficiency on polarized electron transport in monolayer semiconductors possessing spin-valley locking.*

[*] These authors contributed equally to this work
[†] Correspondence: cerobert@insa-toulouse.fr, xavier.marie@insa-toulouse.fr

Atomically thin Transition Metal Dichalcogenides (TMD) semiconductors such as $MoS_2$ have sparked a renewed interest in exploiting both spin and valley degrees of freedom, owing to the remarkable spin-valley locking effects that originate from their lack of inversion symmetry and strong spin-orbit coupling [1–4]. This dictates that the spin and valley degrees of freedom for carriers in the band extrema (electron in the bottom conduction band or holes in the top valence band) are mutually protected; i.e. relaxation requires both a change of valley (for instance with a momentum-conserving phonon), and also a spin-flip[1]. An important consequence is that electron or hole spin relaxation times can reach very large values – in the microsecond range [5–7]. One can expect that this spin-valley locking will also have a strong impact on the lateral transport of electron spin polarization [8]. However, very little is known about the spatial dependence of free carrier spin polarization in these TMD monolayers (ML) despite its crucial relevance for spin(valley)tronics applications using 2D materials [9–11]. Spatial mapping of lateral spin transport in these 2D layers is also important from the point of view of fundamental physics, as it should reveal critical information about possible magnetic phase transitions and valley-polarized collective states that have been theoretically predicted in electron- or hole-doped TMD ML [12–14].

Spatial mapping studies of spin polarization are scarce since (i) in transport experiments, the fabrication of four-terminal nonlocal geometry devices is very challenging with a single TMD ML [15,16] and (ii) in optical measurements, the properties are usually dominated by robust exciton complexes that are characterized by picosecond lifetimes and very limited diffusion lengths, *i.e.* typically ~1 µm at low temperature [17–25]. Small spin diffusion lengths of holes $L_s$ less than 0.1 µm were estimated from valley Hall effect measurements in a $WSe_2$ ML [26] whereas electron spin transport investigations in few-layer $MoS_2$ using a two-terminal spin-valve configuration geometry yielded $L_s$ ~ 0.4 µm [27]. The spin-valley diffusion properties have also been investigated in $WS_2$-$WSe_2$ bilayer heterostructure where the diffusion length is controlled by both inter-layer excitons and resident holes [28].

Here, we locally polarize the Fermi sea of resident electrons in a n-doped $WSe_2$ ML with a *circularly* polarized pump laser, and study how the imbalance between spin-up (K' valley) and spin-down (K valley) electrons evolves in space by using a weak *linearly* polarized probe laser to induce photoluminescence (PL) at a tunable distance $d$ from the pump spot. The circular polarization of the induced trion PL reveals the spin polarization of the resident electrons at the probe's location. We demonstrate that the polarization of resident electrons can propagate over very large distances; we detect polarizations as large as 25% at a pump-probe separation of $d$=20 µm. The spatial decay of the electron polarization yields typical spin diffusion length up to $L_s$=18 µm at T=7 K. Finally, we show that the spin/valley pumping efficiency decreases as the temperature increases, as a consequence of the strongly temperature-dependent spin/valley relaxation time.

Figure 1a presents a schematic of the pump-probe PL experiment performed on a high-quality charge-adjustable $WSe_2$ ML encapsulated in hBN [29]. Continuous wave He-Ne laser beams ($\lambda$=632.8 nm) are used for both pump and probe. Using the same high numerical aperture objective, the two beams are focused on the sample at two different positions separated by a distance $d$. Spots sizes are ~1 µm diameter. The pump beam is right circularly polarized ($\sigma$+) whereas the probe beam is linearly polarized ($\sigma_x$). We detect both right ($\sigma$+) and left ($\sigma$-) circularly polarized luminescence triggered by the probe beam, as a function of the pump-probe separation $d$. Details of the sample fabrication and the experimental set-up are given in the Supplemental Materials S1 and S3.

We first show in Figure 1b the characteristic PL response of the device as a function of bias voltage (i.e. electron density) in response to a single excitation laser. In agreement with many previous reports, the PL spectra in the high energy range are dominated by the recombination of the bright ($X^0$) the dark ($X^D$) neutral exciton and the well identified intravalley (singlet) $X^{S-}$ and intervalley (triplet) $X^{T-}$ negatively charged excitons which are composed of two electrons and one hole (see Figure 1c) [30–32]. In the moderate doping density regime investigated here (a few $10^{11}$ cm$^{-2}$), the resident electrons only populate the lower conduction bands in both K and K' valleys. Note that in this small doping regime the three particle picture (i.e. trion) and the Fermi-polaron description are both relevant [10,33].

---

[1] In the following of the text and for a sake of simplicity we will mainly use the term spin instead of spin/valley knowing that the two degrees of freedom are coupled.

Importantly the degree of circular polarization of the $X^{T-}$ and $X^{S-}$ PL can serve as a quantitative probe of the spin/valley polarization of the resident electrons, as demonstrated in ref [34] and as summarized below. Assuming that the trions are formed through the binding of photo-generated bright excitons with a resident electron (i.e. a bimolecular formation process [35]), the circular polarization of the triplet $P_c(X^{T-})$ and singlet $P_c(X^{S-})$ are simply related to the spin polarization of the resident electrons $P_e = \frac{n_e^{\uparrow K'} - n_e^{\downarrow K}}{n_e^{\uparrow K'} + n_e^{\downarrow K}}$ (where $n_e^{\uparrow K'}$ and $n_e^{\downarrow K}$ are the populations of resident electrons with spin up (K' valley) and spin down (K valley)) and the polarization of the photo-generated excitons $P_0 = \frac{N_0^K - N_0^{K'}}{N_0^K + N_0^{K'}}$ (where $N_0^K$ and $N_0^{K'}$ are the populations of photo-generated bright excitons with carriers in the K and K' valleys) :

<u>Triplet</u> $$P_c(X^{T-}) = \frac{P_0 + P_e}{1 + P_0 P_e} \quad (1)$$

<u>Singlet</u> $$P_c(X^{S-}) = \frac{P_0 - P_e}{1 - P_0 P_e} \quad (2)$$

By measuring both $P_c(X^{T-})$ and $P_c(X^{S-})$ we can thus easily quantify the polarization of the resident electrons $P_e$.

We first characterize in Figure 2a the PL spectra emitted following a circularly polarized (σ+) laser excitation (power=5 µW and electron doping density=4x10$^{11}$ cm$^{-2}$). The circular polarization of the triplet trion $X^{T-}$ PL reaches very large positive values, while it is negative for the singlet trion $X^{S-}$ PL. This is a direct consequence of a spin/valley pumping mechanism that dynamically polarizes the resident electrons in the K' valley with spin up [34]. Note that the efficiency of the spin/valley pumping mechanism depends on both excitation power and doping density. The electron density chosen here maximizes the electron spin polarization while keeping a sufficiently intense trion PL [34]. Using the values of $P_c(X^{T-})$ and $P_c(X^{S-})$, we find that the polarization of resident electrons induced by the laser attains values as large as $P_e$=76% (i.e. the resident electrons mainly populate the lower spin-up conduction band in the K' valley) while the polarization of the photogenerated excitons $P_0$ is 51%.

We then show, in Figure 2b, the PL spectra emitted following a weak linearly polarized (σ$_X$) laser excitation (power=200 nW). As expected, there is no circular polarization of $X^{T-}$ or $X^{S-}$ PL because $P_0 = 0$ (linear excitation) and $P_e = 0$ (no spin/valley pumping mechanism for linear excitation).

Figure 2c shows the first key result of our work. It presents the PL spectra induced by the weak σ$_X$ laser (probe), but now in the presence of the σ+ laser (pump) that is separated by a distance *d*=15.6 µm. Remarkably, now we observe a very large circular polarization for both trions (~+50% for $X^{T-}$ and ~-40% for $X^{S-}$). Because $P_0 = 0$ for the probe (linear excitation), this result directly reveals the polarization of the resident electrons at the location of the probe (i.e. $P_c(X^{T-}) = P_e$ and $P_c(X^{S-}) = -P_e$ ; see equations (1) and (2)). Because the linearly-polarized probe itself does not polarize the resident electrons (as shown above), this demonstrates that the spin polarization of the resident electrons induced by the pump propagates in the 2D layer plane and can be detected by measuring the circular polarization of trion PL below the probe. A key advantage of this PL-based pump-probe experiment in comparison to well-known and powerful Kerr/Faraday rotation spin imaging methods [36,37] is that it allows to quantify, in absolute terms, the spin polarization of the electron Fermi sea. We show in the Supplemental Materials S5 and S6 the dependence of the signal on both pump and probe power as well as the dependence with the doping density.

Figure 3 presents the main result of this work. It displays the dependence of the probe PL circular polarization of both trions as a function of the pump-probe separation *d*. A spatial decay of the spin polarization of resident electrons is clearly evidenced, but the resident electron polarization induced by the pump can propagate on length scales larger than 20 µm. We find that the spin polarization decays approximately exponentially with a spin diffusion length of $L_s$=18 ± 3 µm. This is among the longest spin diffusion lengths reported in semiconductors, despite a modest carrier mobility [38–40]. It is ten times larger than $L_s$ measured in Silicon or p-type GaAs at low temperature, and is similar to the spin diffusion length measured in n-doped GaAs bulk [37,41–43] or quantum wells [44–46]. Remarkably, the spin diffusion length we measure here for a WSe$_2$ ML is very similar to the one determined in graphene monolayers, which are usually characterized by a much larger electron mobility and lower spin-orbit coupling. Using 'non-local' spin valve geometries in graphene, spin diffusion lengths of 2 µm were measured and record values of $L_s$=30 µm were more recently reported [47,48]. This underlines the key role played by the spin-valley locking effect in TMD ML on carrier spin propagation. Note that the measured spin diffusion length of 18 µm is consistent with electron spin/valley relaxation time and electron mobilities recently measured in very comparable n-doped WSe$_2$ ML devices. In a simplified picture based on Einstein relations, we can infer a calculated spin diffusion length of $L_s = \sqrt{D_s \tau_s}$ ~10 µm, using:

(i) the electron mobility $\mu_e$ recently measured in hBN encapsulated TMD MLs - typically 3000 cm$^2$/(V.s) [38–40] and assuming that the spin diffusion coefficient is equal to the charge diffusion coefficient ($D_s = D_c \approx \mu_e kT/e$) [46,48];

(ii) the spin/valley relaxation time obtained from time-resolved Kerr rotation (~ 1 µs for a doping density of about 4x10$^{11}$ cm$^{-2}$) shown in Supplementary Materials S7.

Next, we demonstrate that the electron spin polarization detected at the probe location smoothly tracks the helicity of the pump beam as expected for spin diffusion process. Figure 3b presents the circular polarization degree of X$^{T-}$ and X$^{S-}$ at the probe spot when the pump spot is continuously tuned from purely circular σ+ to purely circular σ- through elliptical and linear polarizations. As expected we observe a change of sign of PL circular polarization when the helicity of the pump is reversed, and a near-linear dependence of the electron spin polarization on the circular polarization degree of the pump excitation.

Finally, we investigate the temperature dependence of the lateral transport of electron spin in the WSe$_2$ ML. Figure 4a shows the spin polarization of the resident electrons $P_e$ as a function of the pump-probe separation, and at various temperatures. Because slightly different values of $P_e$ can be inferred from the polarization of X$^{T-}$ and X$^{S-}$ (see Figure 3), we plot here an average between the two values ($P_e = \frac{P_c(X^{T-}) - P_c(X^{S-})}{2}$). While electron spin transport can be clearly observed up to a temperature of 25 K, the amplitude of the spin polarization decreases compared to the measurements at T=5 K. For temperatures larger than 25 K, no spin polarization can be observed at a distance larger than 10 µm. It turns out that the main origin of this drop of the non-local spin polarization is the decrease of the efficiency of the spin pumping itself, *i.e.* the generation of spin polarized resident electrons by the pump. Figure 4b displays temperature dependence of the resident electron spin polarization obtained from the measured X$^{T-}$ and X$^{S-}$ trions PL circular polarization induced by the pump and detected at the pump location [34] (same experiment as in Figure 2a, raw data are shown in Supplemental Materials S8). We observe that the dynamical polarization of the resident electron decreases drastically between 5 and 30 K. This is a consequence of the decrease of the spin/valley pumping efficiency due itself to the decrease of the spin-valley relaxation time which drops by a factor ~10 in this temperature range, as measured recently by time-resolved Kerr rotation experiments on a similar gated and hBN-encapsulated WSe$_2$ ML [6].

We note that our spatially-resolved studies are consistent with lateral diffusion of spin polarized electrons in the WSe$_2$ monolayer, and do not show evidence of a spontaneous magnetic ordering in the spin-polarized electron gas, as was theoretically predicted to occur in TMD MLs as a consequence of strong exchange interactions [12–14,49]. Specifically, the measured non-local electron polarization varies smoothly with changes in pump intensity, pump helicity, and temperature, and does not show any abrupt discontinuities or saturation or hysteresis that would indicate a transition to an ordered ferromagnetic phase. Moreover, the approximately exponential spatial decay of the polarization signal is in line with expectations of a spin diffusion process, and does not show any sudden variations that might be expected from the formation of ferromagnetically-ordered domains

In conclusion we have investigated the spin/valley diffusion transport of free electrons in a WSe$_2$ ML. In contrast to previous investigations in TMD MLs where the diffusion properties were dominated by short-lived exciton complexes, the efficient spin/valley pumping of resident electrons allow us to evidence transport of spin/valley information over very long distances, with a typical diffusion length of 18 µm. This is a consequence of the long spin/valley relaxation time induced by the unique spin-valley locking effect in this atomically-thin crystal. To separate the effects of diffusion and relaxation, spin grating experiments could be performed in the future to measure accurately the spin diffusion coefficient independently from the charge diffusion coefficient [44,45]. We also anticipate that improvements in the charge mobility of carriers in higher quality material [50] may result in longer spin diffusion length. Spatially-extended hole spin diffusion is also expected in these TMD monolayers, given similarly long spin/valley lifetimes and even stronger spin-valley locking due to the huge spin-orbit splitting in the valence bands. The control of the spin transport properties with an in-plane electric field is the next challenge for future possible applications in spin/valley-tronics.

**Acknowledgements**

We thank I. Paradeisanos, F. Cadiz, T. Amand and H. Dery for useful discussions. This work was supported by the French Agence Nationale de la Recherche under the program ESR/EquipEx+ (grant number ANR-21-ESRE-0025 and the ANR projects ATOEMS, Sizmo-2D and Magicvalley. This study has been partially supported through the EUR grant NanoX n° ANR-17-EURE-0009 in the framework of the "Programme des Investissements d'Avenir. The NHMFL is supported by the National Science Foundation DMR-1644779, the State of Florida, and the US Department of Energy.

**Figures**

Figure 1:

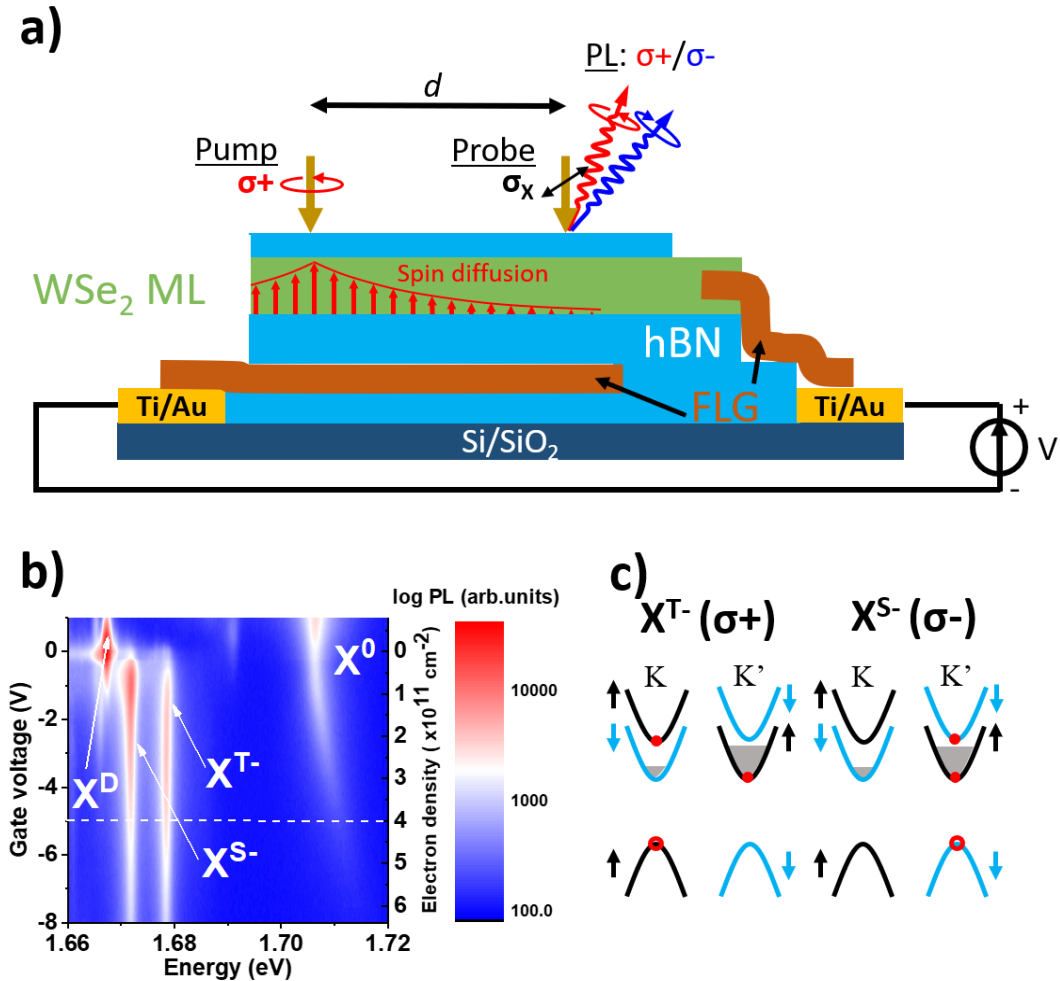

*Figure 1 : a) Sketch of the charge tunable WSe$_2$ ML (not to scale). Two laser spots (pump and probe) separated by a distance d, are focused on the sample. The pump is circularly polarized (σ+) and dynamically polarizes the resident electrons in the K' valley with spin up. This spin/valley polarization diffuses over long distances (sketched by the vertical red arrows representing the average electron spin along the direction perpendicular to the ML) and is detected by a linearly polarized (σ$_X$) probe. The circular polarization of the probe-induced X$^{S-}$ and X$^{T-}$ PL provides a quantitative measurement of the polarization of the 2D electron sea at the location of the probe spot. b) Characterizing the PL from the sample vs. gate voltage (i.e. electron doping density) for the case of a single excitation laser. The horizontal dashed line indicates the doping region where the experiment is conducted. c) Three particle configurations of triplet (X$^{T-}$) and singlet (X$^{S-}$) trions, when resident electrons are polarized in the K' valley. X$^{T-}$ (X$^{S-}$) mainly emits σ+ (σ-) PL.*

Figure 2:

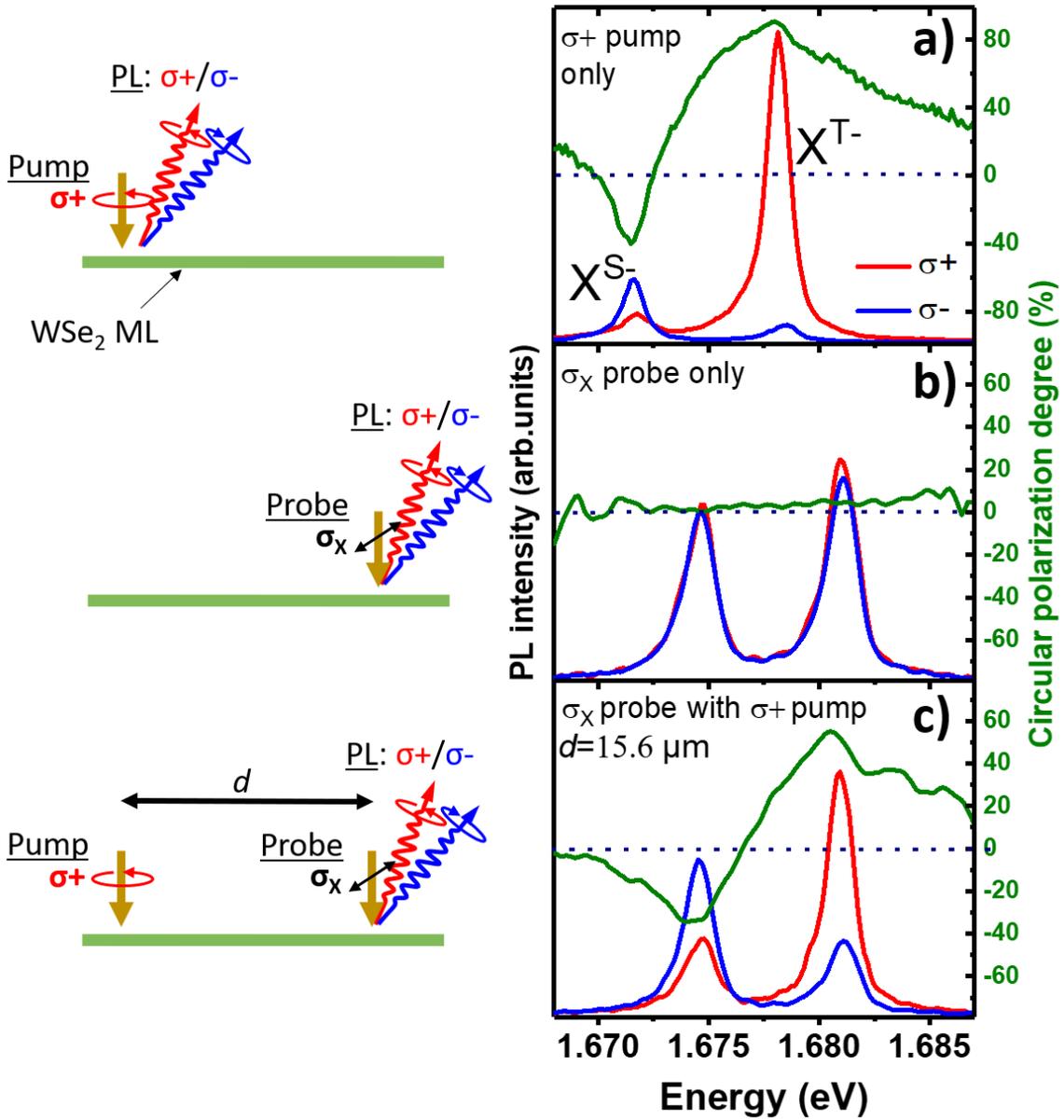

*Figure 2 : Right (σ+) and left (σ-) circularly-polarized PL spectra and corresponding circular polarization degree, in response to a) the σ+ pump only, b) the linear $\sigma_X$ probe only and c) the $\sigma_X$ probe in the presence of a σ+ pump separated by distance d=15.6 µm. The strong circular polarization of the trion PL demonstrates that the resident electrons are strongly spin/valley polarized at the position of the probe laser. T=5 K.*

Figure 3:

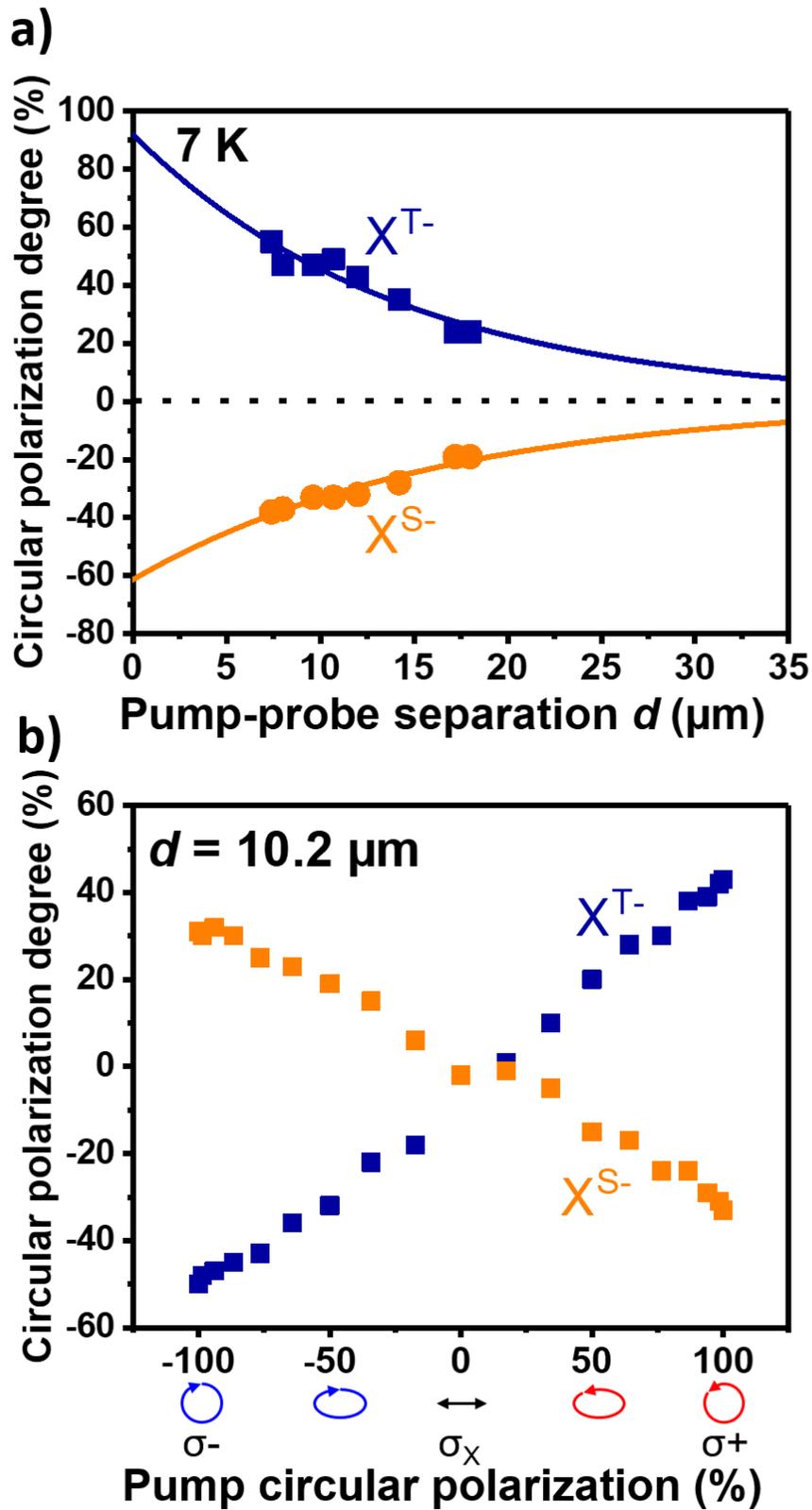

*Figure 3: a) Circular polarization degree of the $X^{T-}$ and $X^{S-}$ trion PL, as a function of the pump-probe separation d. The solid lines are exponential fits. b) Circular polarization degree of $X^{T-}$ and $X^{S-}$ trion PL, at a fixed pump-probe separation, versus the pump helicity. T=7 K.*

Figure 4:

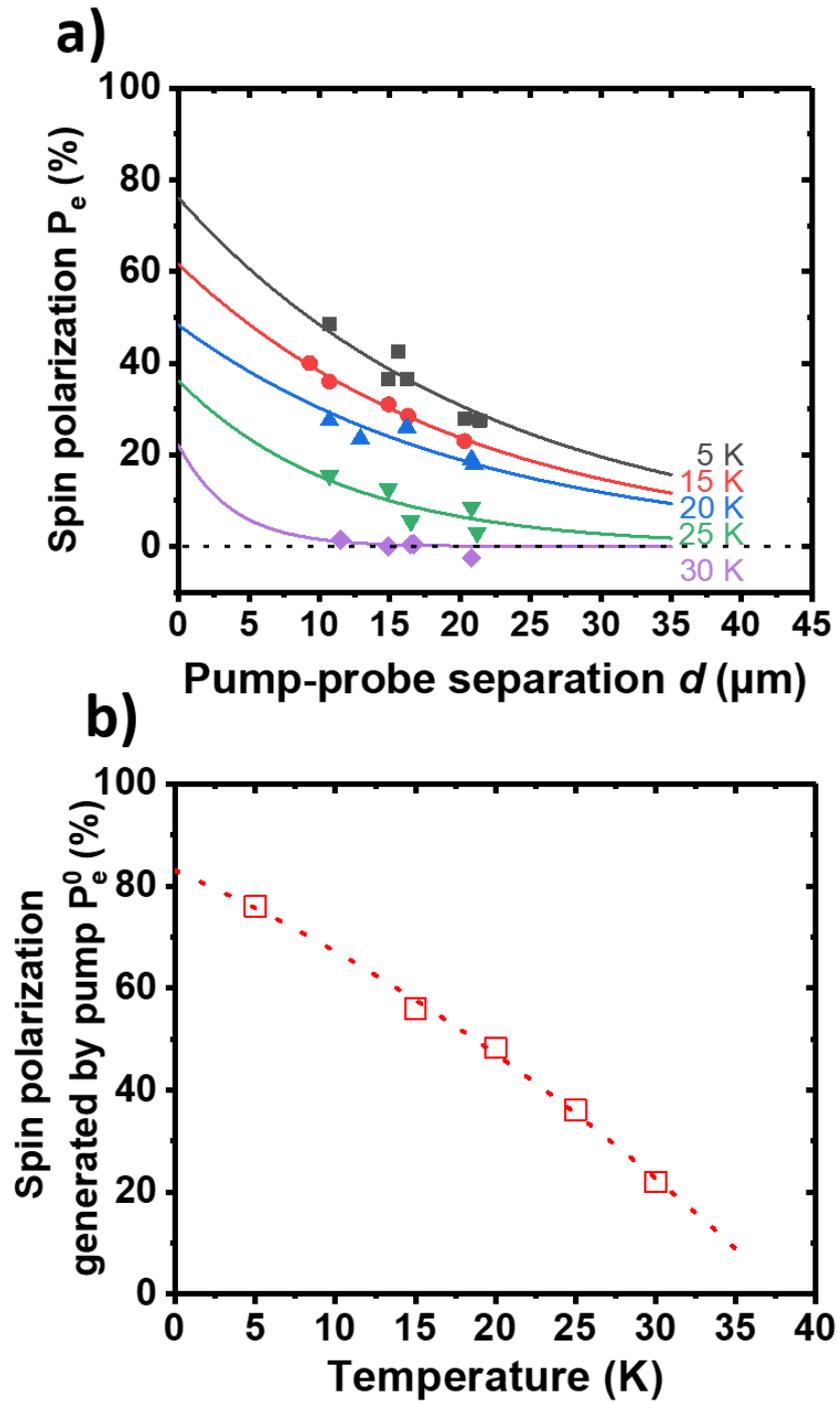

*Figure 4 : a) Temperature dependence of the spin polarization of the resident electrons ($P_e$) at the probe location, versus pump-probe separation d. b) Spin polarization generated below the pump spot ($P_e^0$) as a function of temperature, as extracted from the measurement of the circular polarization of $X^{T-}$ and $X^{S-}$ shown in Supplementary Materials S8. The red dotted line is a guide for the eye. These values of $P_e^0$ are used to fit the results of panel a with monoexponential decays $P_e(d) = P_e^0 exp(-d/L_S)$ that are shown by the solid lines.*

# References

[1]     D. Xiao, G.-B. Liu, W. Feng, X. Xu, and W. Yao, *Coupled Spin and Valley Physics in Monolayers of MoS$_2$ and Other Group-VI Dichalcogenides*, Phys. Rev. Lett. **108**, 196802 (2012).

[2]     G. Sallen, L. Bouet, X. Marie, G. Wang, C. R. Zhu, W. P. Han, Y. Lu, P. H. Tan, T. Amand, B. L. Liu, and B. Urbaszek, *Robust Optical Emission Polarization in MoS$_2$ Monolayers through Selective Valley Excitation*, Phys. Rev. B **86**, 081301 (2012).

[3]     K. F. Mak, K. He, J. Shan, and T. F. Heinz, *Control of Valley Polarization in Monolayer MoS2 by Optical Helicity*, Nat. Nano. **7**, 494 (2012).

[4]     T. Cao, G. Wang, W. Han, H. Ye, C. Zhu, J. Shi, Q. Niu, P. Tan, E. Wang, B. Liu, and J. Feng, *Valley-Selective Circular Dichroism of Monolayer Molybdenum Disulphide*, Nat. Comm. **3**, 887 (2012).

[5]     P. Dey, L. Yang, C. Robert, G. Wang, B. Urbaszek, X. Marie, and S. A. Crooker, *Gate-Controlled Spin-Valley Locking of Resident Carriers in WSe$_2$ Monolayers*, Phys. Rev. Lett. **119**, 137401 (2017).

[6]     J. Li, M. Goryca, K. Yumigeta, H. Li, S. Tongay, and S. A. Crooker, *Valley Relaxation of Resident Electrons and Holes in a Monolayer Semiconductor: Dependence on Carrier Density and the Role of Substrate-Induced Disorder*, Phys. Rev. Materials **5**, 044001 (2021).

[7]     J. Kim, C. Jin, B. Chen, H. Cai, T. Zhao, P. Lee, S. Kahn, K. Watanabe, T. Taniguchi, S. Tongay, M. F. Crommie, and F. Wang, *Observation of Ultralong Valley Lifetime in WSe2/MoS2 Heterostructures*, Science Advances **3**, e1700518 (2017).

[8]     L. Wang and M. W. Wu, *Electron Spin Diffusion in Monolayer MoS$_2$*, Phys. Rev. B **89**, 205401 (2014).

[9]     X. Lin, W. Yang, K. L. Wang, and W. Zhao, *Two-Dimensional Spintronics for Low-Power Electronics*, Nat. Elec. **2**, 10 (2019).

[10]    P. Back, M. Sidler, O. Cotlet, A. Srivastava, N. Takemura, M. Kroner, and A. Imamoğlu, *Giant Paramagnetism-Induced Valley Polarization of Electrons in Charge-Tunable Monolayer MoSe$_2$*, Phys. Rev. Lett. **118**, 237404 (2017).

[11]    M. E. Flatté and J. M. Byers, *Spin Diffusion in Semiconductors*, Phys. Rev. Lett. **84**, 4220 (2000).

[12]    J. E. H. Braz, B. Amorim, and E. V. Castro, *Valley-Polarized Magnetic State in Hole-Doped Monolayers of Transition-Metal Dichalcogenides*, Phys. Rev. B **98**, 161406 (2018).

[13]    D. K. Mukherjee, A. Kundu, and H. A. Fertig, *Spin Response and Collective Modes in Simple Metal Dichalcogenides*, Phys. Rev. B **98**, 184413 (2018).

[14]    J. G. Roch, Dmitry Miserev, G. Froehlicher, N. Leisgang, L. Sponfeldner, K. Watanabe, T. Taniguchi, J. Klinovaja, D. Loss, and R. J. Warburton, *First-Order Magnetic Phase Transition of Mobile Electrons in Monolayer MoS$_2$*, Phys. Rev. Lett. **124**, 187602 (2020).

[15]    T. S. Ghiasi, J. Ingla-Aynés, A. A. Kaverzin, and B. J. van Wees, *Large Proximity-Induced Spin Lifetime Anisotropy in Transition-Metal Dichalcogenide/Graphene Heterostructures*, Nano Lett. **17**, 7528 (2017).

[16]    Y. K. Luo, J. Xu, T. Zhu, G. Wu, E. J. McCormick, W. Zhan, M. R. Neupane, and R. K. Kawakami, *Opto-Valleytronic Spin Injection in Monolayer MoS$_2$/Few-Layer Graphene Hybrid Spin Valves*, Nano Lett. **17**, 3877 (2017).

[17]    M. Schwemmer, P. Nagler, A. Hanninger, C. Schüller, and T. Korn, *Long-Lived Spin Polarization in n-Doped MoSe$_2$ Monolayers*, Appl. Phys. Lett. **111**, 082404 (2017).

# Supplemental Material for "Optical detection of long electron spin transport lengths in a monolayer semiconductor"

L. Ren[1], L. Lombez[1], C. Robert[1], D. Beret[1], D. Lagarde[1], B. Urbaszek[1], P. Renucci[1], T. Taniguchi[2], K. Watanabe[3], S.A. Crooker[4], X. Marie[1]

[1]Université de Toulouse, INSA-CNRS-UPS, LPCNO, 135 Av. Rangueil, 31077 Toulouse, France
[2]International Center for Materials Nanoarchitectonics, National Institute for Materials Science, 1-1 Namiki, Tsukuba 305-00044, Japan
[3]Research Center for Functional Materials, National Institute for Materials Science, 1-1 Namiki, Tsukuba 305-00044, Japan
[4]National High Magnetic Field Laboratory, Los Alamos National Laboratory, Los Alamos, New Mexico 87545, USA

## S1. Sample fabrication

A van der Waals heterostructure made of an exfoliated ML-WSe$_2$ embedded in high quality hBN crystals [1] was fabricated by using a dry stamping technique [2] in the inert atmosphere of a glove box. The exfoliated layers were precisely transferred onto a SiO$_2$/Si substrate with pre-patterned Ti/Au electrodes. Flux-grown bulk crystals of WSe$_2$ were purchased from 2D semiconductors. Flakes of few-layer of graphene (FLG) were exfoliated from a HOPG bulk crystal and were used for the back gate and to contact the ML-WSe$_2$. A detailed sketch and a microscope image of the sample is shown in Figure S1.

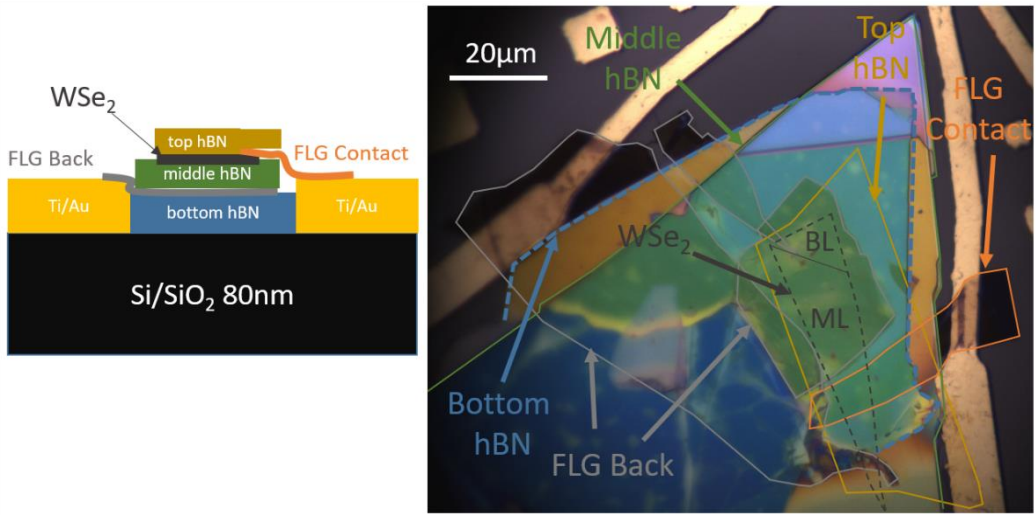

*Figure S1 : Sketch and optical microscope image of the WSe$_2$ charge tunable device.*

## S2. Evaluation of the carrier density

The carrier density in the charge tunable device was estimated by two methods, as in our previous work [3,4]. In the first method, the simple plate capacitance model was used to evaluate the carrier density. Given an applied gate voltage ($V$), the middle hBN thickness (210 nm in our sample) and using the hBN dielectric constant of $\varepsilon_{hBN} \sim 3$ [5,6], the variation of electron density $\Delta n$ is related to the variation of the applied gate voltage $\Delta V$ through $\Delta n = \frac{\varepsilon_0 \varepsilon_{hBN}}{e \times t} \Delta V$. Alternatively, we used the oscillations in the reflectivity spectrum of the bright exciton as a function of gate voltage in the p-doped regime observed at +9 T. As demonstrated in Ref [7], these oscillations are due to the interaction of the exciton with the quantized Landau levels of the hole Fermi sea. The period of the oscillations $\Delta V_{LL}$ is related to the filling of one Landau level $P_{LL} = \frac{eB}{2\pi\hbar} = 2.18 \times 10^{11}$ cm$^{-2}$. We can thus calculate the hole density as a function of the gate voltage by: $\Delta p = \Delta V \frac{P_{LL}}{\Delta V_{LL}}$. This yielded the same estimation of the carrier density as the one deduced from the capacitance model. The advantage of this method is that it does not require knowledge of material parameters.

**S3. Experimental setup**

The experimental setup is sketched in Figure S2. The sample and a high numerical aperture objective (NA=0.82) are placed in a vibration-free closed-cycle He cryostat. The sample is moved with x-y-z piezo-positioners. Two HeNe laser beams (pump and probe) are directed towards the sample through mirrors (M) and beam splitters (BS and BS1). Mirrors (M) and beam splitters (BS) are placed on kinematic mirror mounts to allow adjusting the distance between pump and probe spots on the sample. The distance between the two spots is determined through an imaging system (not sketched). The polarizations of the pump and probe lasers were adjusted with a combination of linear polarizer, half wave plate (λ/2) and quarter wave plate (λ/4). We rotated the λ/2 and λ/4 plates to pre-compensate the polarization before the mirrors and beam splitters, to ensure that the pump and probe beams were circularly (σ+) polarized and linearly polarized, respectively, before entering the cryostat. Unless stated otherwise, we fixed the pump and probe excitation powers to 5 µW and 200 nW, respectively. Spot diameters were estimated to be around 1 µm diameter. The photoluminescence signal was collected by the same microscope objective and was directed to a spectrometer equipped with a Si-CCD camera. In the detection path we placed a quarter wave plate (λ/4) and a linear polarizer. By rotating the λ/4 plate we selected either σ+ or σ- PL signal. Note that we used a second beam splitter (BS2) identical to BS1 as a polarization compensator. We adjusted the alignment to optimize the detection of the PL induced by the probe. Nevertheless, we could still detect a significant part of the PL induced by the pump, especially for small pump-probe separations. To mitigate this issue, we used a subtraction protocol that is presented in the next section.

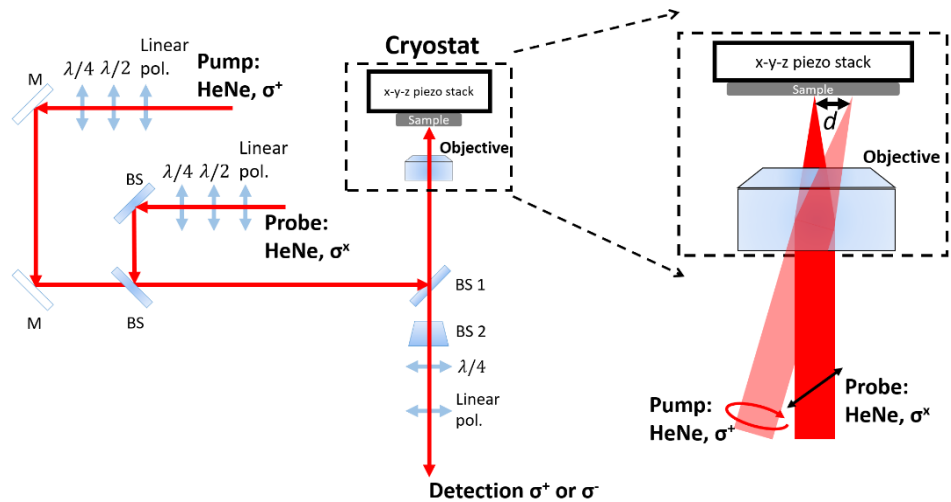

*Figure S2 : Sketch of the experimental setup.*

**S4. Protocol of measurement and data processing**

In order to eliminate the PL signal coming from the pump, we first measured the PL signal (σ+ and σ- detections) in the presence of both pump and probe, named $I^{\sigma+}_{pump-probe}$ and $I^{\sigma-}_{pump-probe}$. Then we performed the measurement only with the pump and recorded again σ+ and σ- at the detection. We name these signals $I^{\sigma+}_{pump}$ and $I^{\sigma-}_{pump}$. We obtained the net σ+ and

σ- PL signals just under the probe position (shown in Figure 2c of the main text) by $(I^{\sigma+}_{pump-probe} - I^{\sigma+}_{pump})$ and $(I^{\sigma-}_{pump-probe} - I^{\sigma-}_{pump})$. The circular polarization is defined as:

$$P_c = \frac{\left(I^{\sigma+}_{pump-probe}-I^{\sigma+}_{pump}\right)-(I^{\sigma-}_{pump-probe}-I^{\sigma-}_{pump})}{\left(I^{\sigma+}_{pump-probe}-I^{\sigma+}_{pump}\right)+(I^{\sigma-}_{pump-probe}-I^{\sigma-}_{pump})}.$$

Note that despite this protocol it is difficult to obtain reliable results for pump-probe separations less than ~8 µm because the PL signal due to the pump laser is much larger than the signal due to the probe. Also, our scan range is limited to ~25 µm because of the limited aperture of the objective.

### S5. Dependence on pump and probe power

We present in Figure S3 the circular polarization of $X^{T-}$ and $X^{S-}$ PL in the pump-probe experiment as a function of both pump and probe power, for a fixed pump-probe separation *d* and at T=5 K. As expected, the circular polarization drops smoothly when the pump power decreases. This is due to the decrease of the spin/valley pumping efficiency (i.e. there are not enough photogenerated carriers to spin/valley polarize the resident electrons). Opposite behavior is observed for the probe power dependence. When the probe power is too large, the photoexcited carriers generated by the probe tend to depolarize the resident electrons. In the main text we present the results using a very small probe power of 200 nW, a value that minimizes the effect of depolarization by the probe while keeping measurable PL signal.

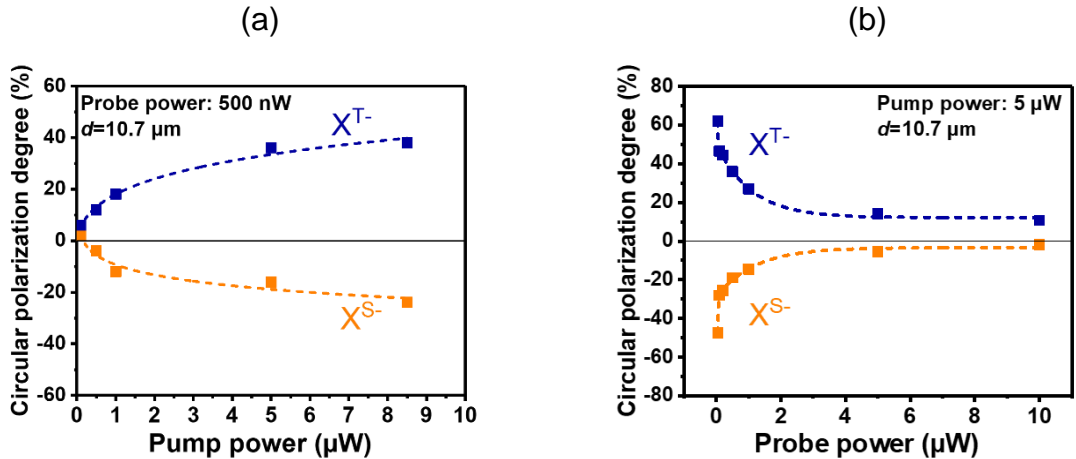

*Figure S3 : PL circular polarization of $X^{T-}$ and $X^{S-}$ induced by the probe for a fixed pump-probe separation as a function of (a) the pump power and (b) the probe power. Dashed lines are guides for the eye. T=5 K.*

## S6. Dependence on doping density

We present in Figure S4, the PL circular polarization induced by the probe as a function of the electron doping density for a fixed pump-probe separation. We observe that the measured spin/valley polarization is smoothly reduced when we decrease the doping density. Here again, this can be interpreted as a drop of the spin/valley pumping efficiency. Note that in our previous work on the spin/valley pumping mechanism (using only one laser spot) [4] we observed that the circular polarization of $X^{S-}$ following a σ+ excitation switched from negative to positive when we decreased the doping density. In that work, we could not clearly explain this result. The results of Figure S4 show that this is due to a decrease of the spin/valley polarization of the resident electrons.

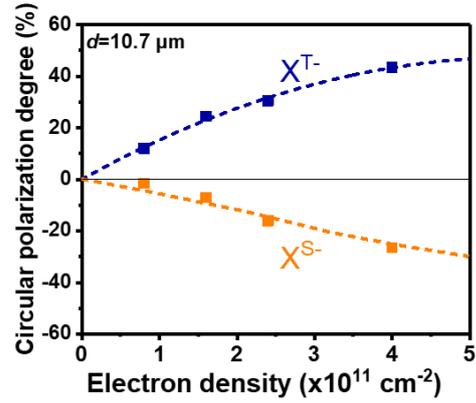

Figure S4 : PL circular polarization of $X^{T-}$ and $X^{S-}$ induced by the probe for a fixed pump-probe separation as a function of the doping density. Pump power: 5 µW, probe power: 200 nW, T=5 K.

## S7. Spin/valley relaxation time

We present in Figure S5, the time-resolved Kerr rotation (TRKR) measurement of resident electron spin lifetimes in an electrostatically-gated hBN/WSe$_2$/hBN sample with very similar layer structure and optical quality in comparison to the gated hBN/WSe$_2$/hBN sample studied in the spin diffusion measurements described in the main text. Spin/valley lifetimes in this sample were reported recently in [8] and details of the ultrafast optical pump-probe experiment can be found therein. The spatially-resolved studies of spin diffusion described in the main text were performed at a background electron density of 4 x 10$^{11}$cm$^{-2}$ (i.e., midway between the two densities at which the TRKR studies shown in Figure S5 were explicitly performed), and at similarly low temperatures. We therefore estimate that the spin-valley lifetime of the resident electrons in the spin diffusion studies is long, on the order of 1 µs.

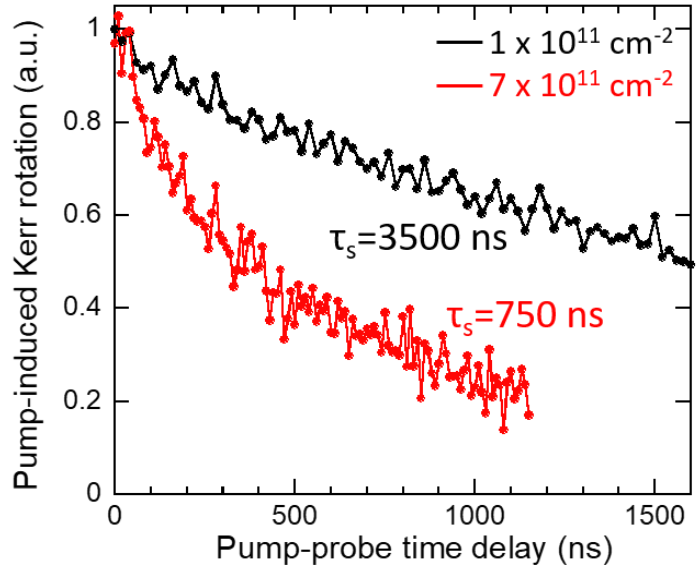

Figure S5 : TRKR measurement of resident electron spin lifetimes, in an electrostatically-gated hBN/WSe$_2$/hBN sample. The black and red curves were acquired at background electron densities of approximately 1 x 10$^{11}$/cm$^2$ and 7 x 10$^{11}$/cm$^2$, and reveal electron spin/valley lifetimes of ~3500 and ~750 ns, respectively, at a temperature of 5.8 K.

## S8. Temperature dependence of the spin/valley pumping mechanism

Figure 4b of the main text presents the spin polarization generated by the pump. We show here the raw data that enable us to extract these values. Figure S6 shows the PL induced by the circular pump only (no probe) at various temperatures. Measuring both circular polarization of the triplet $P_c(X^{T-})$ and singlet $P_c(X^{S-})$ trions and using their relation to the spin polarization of the resident electrons $P_e$ and the polarization of the photo-generated excitons $P_0$ (namely $P_c(X^{T-}) = \frac{P_0+P_e}{1+P_0 P_e}$ and $P_c(X^{S-}) = \frac{P_0-P_e}{1-P_0 P_e}$), we extract $P_e$ and $P_0$ as a function of temperature (results are summarized in the table below). Figure 4b of the main text shows $P_e$ as a function of temperature.

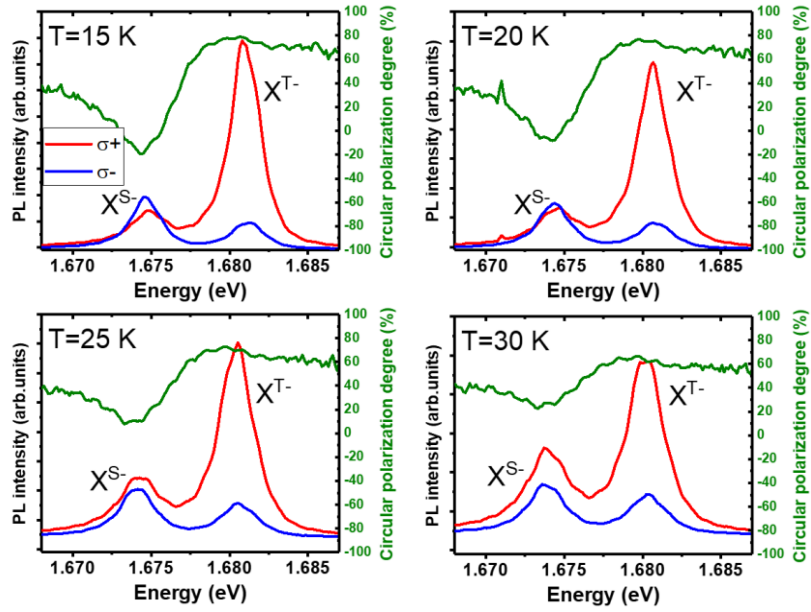

*Figure S6 : PL and circular polarization degree induced by the pump only (power 5 µW) as a function of temperature.*

| T (K) | $P_c(X^{T-})$ | $P_c(X^{S-})$ | $P_0$ | $P_e$ |
|---|---|---|---|---|
| 5  | 91% | -40% | 51% | **76%** |
| 15 | 79% | -19% | 41% | **56%** |
| 20 | 75% | -8%  | 42% | **78%** |
| 25 | 70% | 11%  | 45% | **36%** |
| 30 | 62% | 27%  | 46% | **22%** |

## S9. Additional discussion regarding the difference of polarization between X<sup>T-</sup> and X<sup>S-</sup>

We noted in the main text that the circular polarization of the singlet trion X<sup>S-</sup> PL at the probe location is slightly smaller in absolute value than the circular polarization of the triplet trion X<sup>T-</sup> PL. This is surprising as according to our simple model $P_c(X^{T-})$ should be equal to $P_e$ and $P_c(X^{S-})$ to $-P_e$. We elaborate here a possible explanation of this observation.

The degree of circular polarization of a trion in our cw PL experiment is given by:

$$P = \frac{P_G}{1 + \frac{\tau}{\tau_s}}$$

where $P_G$ is the degree of circular polarization at the generation, $\tau$ is the lifetime of the trion and $\tau_s$ is its spin (or valley) relaxation time (assuming here a single relaxation mechanism). In the simple model presented in the main text, we considered that the measured degrees of circular polarization for the triplet and the singlet directly reflect $P_G$ (the degree of circular polarization generated by the probe spot). In other words, we disregard the role of $\tau$ and $\tau_s$. This was justified in our previous work [4] as $\tau \ll \tau_s$. Nevertheless, we also measured that the lifetime of $X^{S-}$ is longer than the lifetime of $X^{T-}$ in agreement with a slightly stronger oscillator strength for the singlet [9]. Thus, the measured polarization for $X^{S-}$ could be slightly reduced if its lifetime is not completely negligible with respect to its spin/valley relaxation time.